\documentclass[12pt]{article}

\usepackage{scicite}

\usepackage{times}
\usepackage{graphicx} 
\usepackage{amsmath} 
\usepackage{amsfonts} 
\usepackage{caption} 
\usepackage[draft]{hyperref} 

\newenvironment{sciabstract}{%
\begin{quote} \bf}
{\end{quote}}

\title{Broadband localization of light at the termination of a topological photonic waveguide} 

\author
{\parbox{\linewidth}{\centering 
Daniel Muis,$^{1\dag}$ Yandong Li,$^{2\dag}$ Ren\'e Barczyk,$^{3}$ Sonakshi Arora,$^{1}$  L. Kuipers,$^{1\ast}$ Gennady Shvets,$^{2}$ Ewold Verhagen$^{3\ast}$}\\
\\
\parbox{\linewidth}{\centering 
\normalsize{$^{1}$Kavli Institute of Nanoscience, Department of Quantum Nanoscience, Delft  University of Technology, 2628CJ Delft.}\\
\normalsize{$^{2}$School of Applied and Engineering Physics, Cornell University, Ithaca, New York 14853, USA.}\\
\normalsize{$^{3}$Center for Nanophotonics, AMOLF, Science Park 104, 1098XG Amsterdam, the Netherlands.}}\\
\\
\parbox{\linewidth}{\centering 
\normalsize{$^\dag$These authors contributed equally to this work.}\\
\normalsize{$^\ast$Corresponding authors. E-mail: l.kuipers@tudelft.nl, e.verhagen@amolf.nl}
}}

\date{}

\begin{document} 


\baselineskip24pt


\maketitle 


\begin{sciabstract}
  Localized optical field enhancement enables strong light-matter interactions necessary for efficient manipulation and sensing of light. Specifically, tunable broadband energy localization in nanoscale hotspots offers a wide range of applications in nanophotonics and quantum optics. We experimentally demonstrate a novel principle for the local enhancement of electromagnetic fields, based on strong suppression of backscattering. This is achieved at a designed termination of a topologically non-trivial waveguide that nearly preserves the valley degree of freedom. The symmetry origin of the valley degree of freedom prevents edge states to undergo intervalley scattering at waveguide discontinuities that obey the symmetry of the crystal. Using near-field microscopy, we reveal that this can lead to strong confinement of light at the termination of a topological photonic waveguide, even without breaking time-reversal symmetry. We emphasize the importance of symmetry conservation by comparing different waveguide termination geometries, confirming that the origin of suppressed backscattering lies with the near-conservation of the valley degree of freedom, and show the broad bandwidth of the effect. 
\end{sciabstract}

\clearpage
\section*{Introduction}
Manipulation of light in topological photonic platforms offers an exciting paradigm for exploring fundamental principles of light-matter interactions and developing novel communication technology components with new or improved performance\cite{Lu2014,Ozawa2019}. Specifically, photonic systems with non-trivial topology of their bulk eigenstates feature useful phenomena such as helical edge states \cite{Khanikaev2013,Wu2015,Ma2016,Ma2017,Dong2017,Gao2017,Barik2018,parappurath2020}. Careful tailoring of these photonic systems offers opportunities for efficient and robust molding of light on a nanophotonic chip, important for, e.g., coupling of quantum emitters \cite{Lodahl2017}, integrated photonic circuits \cite{Wu2017,Bogaerts2020}, and microcavity lasing \cite{Han2019,zeng2020,yang2022}.

Topological photonic edge states can emerge at an interface of dielectric photonic crystals (PhCs) with broken spatial symmetry \cite{Wu2015,Ma2016}. Such broken symmetry can open a broad band gap in which bidirectional spin- or valley-locked chiral edge states exist. These edge states are known to exhibit suppressed backscattering by spin- or valley-flipping processes at corners that obey the protecting symmetry of the crystal\cite{Barik2018,Arora2021}. This notion of reduced backscattering and robust transmission raises the intriguing question of what happens to optical energy in a system if a topological waveguide is abruptly terminated. Related to this question, in nonreciprocal waveguides it was proposed that optical intensity could be enhanced at a waveguide termination, specifically in waveguides based on magneto-optic surface plasmon-polaritons\cite{Chettiar2014,Shen2015,Tsakmakidis2017,Gangaraj2019,Fernandes2019,Mann2019,Buddhiraju2020} and on bi-anisotropic metamaterials\cite{Fernandes2019}. These envisioned effects have been theorized to have topological origin~\cite{Fernandes2019} and it has been argued that their large and broadband field enhancements can benefit light-matter interactions and nonlinear optics\cite{Mann2021}. This suggests that reciprocal topological waveguides could also facilitate such enhancement, potentially broadening the scope of applications. Indeed, it was predicted that an abruptly terminated photonic valley Hall channel can localize light in a hotspot at the termination~\cite{Li2020}. This topology-enabled buildup of optical energy is related to the near-conservation of the valley degree of freedom, as the minimization of intervalley overlap delays backreflection.

In this work, we experimentally demonstrate localization and enhancement of optical energy at a termination of a topological waveguide at telecom frequencies, using near-field microscopy \cite{Rotenberg2014}. We observe that light intensity is enhanced by an order of magnitude for a broad range of frequencies where there exist no other surface modes that can carry energy away from the termination along the interface terminating the topological waveguide. We reveal that the symmetry of the termination plays an important role in the observed effect, confirming the responsible mechanism of backscattering reduction. These results thus demonstrate a new mechanism for electromagnetic field enhancement at the nanoscale, with consequently broad application potential in the control of light-matter interactions and improved manipulation and sensing of light in photonic technology.

The topological waveguide is designed by connecting two mirror-inverted valley photonic crystals (VPCs)~\cite{Ma2016}. The VPCs have a rhombic unit cell that is composed of two opposite-facing triangular air holes with an inequivalence in size which breaks inversion symmetry and therefore opens a two-dimensional photonic band gap \cite{Ma2016,Dong2017,Chen2017}. The inversion symmetry breaking results in an opposite-signed Berry curvature locally concentrated at the K/K’ valleys in the Brillouin zone. Connecting two VPCs that have distinct topologies, a valley-dependent edge state emerges which emulates the quantum valley Hall effect and is protected against intervalley scattering at defects that conserve the $C_3$ crystalline symmetry~\cite{Gao2017,Arora2021}. Reciprocity guarantees that two edge states associated with opposite valley degree of freedom propagate in opposite directions. As long as the edge state dispersion lies below the light line, light is confined to the dielectric photonic crystal slab.

The topological waveguide is abruptly terminated by an insulating trivial PhC with a two-dimensional photonic band gap that prohibits further propagation of the edge state. This termination results in a triple-junction of three inequivalent interfaces as shown in Fig.~\ref{fig1}. The two interfaces between the VPCs and the PhC can in principle support trivial surface states. These states are not expected to span the entire VPC gap, as the difference in the valley Chern numbers of the VPC and the PhC is 1/2. Instead, each interface exhibits a frequency gap where no propagating states exist. We consider of particular interest the regime where the frequency gaps of both interfaces overlap. This frequency regime is bound by surface modes of both interfaces, and we hence refer to it as the surface mode gap. Operating at a frequency in this gap, light that is incident from the valley Hall waveguide towards the termination can only scatter out-of-plane (to free space radiation), or back into the topological waveguide. However, backscattering requires a flip of the valley index, which is suppressed at a suitably designed termination~\cite{Li2020}. We investigate the resulting signature of suppressed backscattering on broadband optical enhancement at the termination.

\section*{Results}
\subsection*{Experimental observation of optical energy enhancement}
To observe the spatial distribution of light near the termination of a topological waveguide we use scanning-probe near-field microscopy, as shown schematically in Fig.~\ref{fig1}. A heterodyne detection technique allows for phase-resolved detection with high signal-to-noise ratio (see Materials and Methods). Infrared light enters the VPC-VPC interface via a silicon ridge waveguide from the side of a silicon-on-insulator (SOI) sample. As the forward propagating edge state (at the K' valley in k-space, dark orange arrow in Fig.~\ref{fig1}) encounters the termination, it can couple to one of the trivial surface modes that propagates along the VPC-PhC interface (output ports labelled 1 or 2). A near-field probe that is  raster-scanned above the surface picks up a fraction of the evanescent in-plane electric field components with sub-wavelength resolution, limited by the 194~nm aperture of the aluminium-coated near-field probe. The raster scan allows imaging the intensity of light in the photonic crystal as a function of position in two-dimensional space at a height of approximately 20 nm above the surface. The intensity is derived directly from the electric field components and is shown in real space for three different laser frequencies in Fig.~\ref{fig2}A. At a laser frequency of 202.7~THz, the edge state is seen to propagate through the topological waveguide until it terminates upon encountering the trivial PhC. At the termination, the edge state strongly localizes with enhancement of the local field energy in a spot with a spatial extent of $\sim$520~nm. 

The two smaller panels on the right side of Fig.~\ref{fig2}A depict scans from the same area, but at laser frequencies of 194.2 and 202.0 THz. For reasons that are explained below, these images are obtained on samples with slightly different design, distinguished by a small shift of the holes in the PhC unit cell with respect to the VPC crystals (effectively shifting the PhC lattice). We quantify this shift by a parameter called `dislocation' (dl), see Supplementary Materials for the detailed definition. At these frequencies, we observe that an edge state propagates along the VPC-PhC interface away from the termination. The light propagates into either port (1) at 194.2~THz or port (2) at 202.0~THz, which correspond to an acute or obtuse angle, respectively, with respect to the direction of the incident VPC-VPC edge state (see Supplementary Materials for extended plots over the entire sample area). The two characteristic phenomena, namely optical energy localization and the out-coupling to one of the trivial surface modes, agree with simulation results (Fig.~\ref{fig2}B).

Figures~\ref{fig2}A and \ref{fig2}B show that the behaviour of scattering at the termination, as well as the observed localization, depends pronouncedly on the optical frequency. The frequency regimes associated with the different phenomena can be understood from the COMSOL-simulated dispersion diagram shown in Fig.~\ref{fig2}C. The diagram depicts the frequency dependence of the guided mode wavevectors along the direction of propagation in the incident VPC-VPC edge (orange line) and in both output ports (1) and (2) (green and blue lines, respectively). Wavevectors are shown in the range $[\pi/(2\text{a}_0),\pi/ \text{a}_0]$, which is equivalent to the positive part of the first Brillouin zone that lies below the light line. Within the topological bandgap but out of the surface mode gap, incoming light from the valley edge mode can be carried away from the termination in port (1) at low frequencies and in port (2) at high frequencies. When operating at frequencies in the surface mode gap, no propagating surface states are supported by output ports (1,2) while the VPC-VPC interface does support the topological edge state.

Experimentally, we probe the frequency dependence by executing raster scans of the near-field probe for a range of laser frequencies. By employing a two-dimensional Fourier transform of a region containing the VPC-PhC interface and projecting the intensity normal to the direction of the interface, we retrieve the wavevectors of light propagating along the VPC-PhC interface, i.e. along output ports 1 and 2. The results are experimental dispersion diagrams (Fig.~\ref{fig2}D), showing a $\sim$12~THz window that is accessible with our tunable laser. The two panels are taken on structures with different dislocation values of 0.14a$_0$ and 0.06a$_0$. At frequencies below the surface mode gap, a surface mode propagating along port (1) exists at the terminating interface with dl=0.14a$_0$ (Fig. 2D, left panel). For a smaller dislocation, dl=0.06a$_0$, all surface mode dispersion curves shift to lower frequencies, and a surface mode propagating along port (2) emerges (Fig. 2D, right panel). This overall downshift of all surface mode frequencies is because a slight dislocation, which corresponds to a broader spatial gap between the VPC and the PhC lattices, increases the effective index of guided waves along the terminating interface. We note that, for dl=0.06a$_0$, the surface mode along port (1) is shifted to such low frequencies that it overlaps with the bulk modes of the VPC. Thus, both surface modes along the output ports labeled in Figures 1 and 2C can be observed, with their precise frequency tunable through the dislocation parameter. The dispersion diagrams display an offset in frequency of $\sim$7~THz compared to simulation, likely due to slight geometrical deviations of the fabricated sample, but the results are otherwise consistent.
Importantly, from the two dispersion diagrams we retrieve for dislocations 0.14a$_0$ and 0.06a$_0$, it becomes clear that a broadband surface mode gap exists where the incident VPC-VPC edge state cannot couple to either upper or lower propagating surface states. Instead, the edge state may either scatter back, or in the out-of-plane direction to free-space radiation resulting in a loss of the on-chip optical energy. Backscattering is, however, symmetry-dependent and can be suppressed by a termination with particular orientation. This suppression results in an accumulation of optical field energy that we observe in Fig.~\ref{fig2}A.

\subsection*{Role of symmetry in protection and field enhancement}
When a valley edge mode encounters the termination, its evanescent tails penetrate into the two VPC bulks and are affected by the symmetry of the termination. As this can affect the rate of backscattering, we investigate the correlation between the observed localization and the symmetry of the termination. Along terminations with different geometries, the overlap between the forward propagating (K'-valley) and the backscattered (K-valley) modes are significantly different. Some unique choices of the termination can greatly suppress this overlap and hence reduce the rate of backscattering (see Supplementary Materials for detailed theoretical derivation). We distinguish between zigzag terminations, corresponding to $\pi/3$ or $2\pi/3$ angles between the VPC-PhC and VPC-VPC interfaces, and armchair terminations, corresponding to $\pi/2$ angle. We reveal that zigzag terminations largely suppress the overlap, while armchair terminations feature significant overlap. This mathematical conclusion can be intuitively understood as following: a zigzag termination maximally follows the symmetry of the VPC lattice and thus conserves the valley DOF to a great degree, exploiting the topologically non-trivial nature of the VPC crystals. In contrast, an armchair termination strongly breaks the crystal symmetry that protects the VPC-VPC states from scattering efficiently. While a certain degree of symmetry breaking is unavoidable for any shape or termination, and light can thus eventually scatter back, the rate at which this happens thus depends strongly on the symmetry of the termination in the topological crystal\cite{Li2020}. Finally, we note that the finite size of the termination in the three-dimensional slab setting means that some scattering to free space is possible. There is in principle no reason to assume that the amount of free-space scattering as a result of this defect is significantly different for zigzag and armchair terminations.

To study the effect of termination geometry experimentally, we probe the electromagnetic near-field intensity near the end of the waveguide for samples with either zigzag or armchair terminations. First, we experimentally confirm the optical energy localization at a zigzag termination. The green curve in Fig.~\ref{fig3}A shows the intensity as a function of position along the VPC-VPC waveguide, which we obtain by integrating the intensity in a 2D scan over a transversal width of two lattice constants across the center of the waveguide. We observe for both the zigzag termination in Fig.~\ref{fig3}A and the armchair termination in Fig.~\ref{fig3}B an oscillatory behaviour in intensity along the majority of the VPC-VPC waveguide. This is the result of interference between forward and backward propagating modes, for all Bloch harmonics. Our phase-sensitive heterodyne detection scheme allows separating the intensities of forward and backward waves by applying a wavevector filter on the Fourier-transformed fields around the respective Bloch components~\cite{Arora2021}. The inverse Fourier transform using only the $k_x$ components that correspond to the forward (backward) propagating mode then returns the individual contribution to the intensity by that mode, shown by the red (blue) curve in a translationally invariant part of the topological waveguide. We refer to Supplementary Materials section 4 for a detailed explanation of the formalism behind the forward/backward separation. We observe a significant intensity of backward propagation for both the zigzag and armchair terminations. As the observation frequency resides in the surface mode gap, see Fig.~\ref{fig2}D and Supplementary Materials Fig. S3, there can be no energy propagation along the VPC-PhC interfaces. Light does, however, significantly scatter out-of-plane as a consequence of symmetry breaking at the termination, explaining a reduced intensity of the backward propagating mode. Furthermore, the intensity after being reflected from a zigzag or armchair termination is approximately the same, see Supplementary Materials Fig.~S4, indicating that the out-coupled energy to free space does not depend strongly on the type of termination.

Importantly, the local intensity at the zigzag termination is observed to be 16 times larger than the average intensity of the forward propagating mode in the waveguide, in a sharply localized hot spot. In stark contrast, the armchair termination shows no enhancement, meaning that all energy is immediately reflected back. Simulations confirm that optical energy is strongly enhanced only at the zigzag termination, see Supplementary Materials Fig.~S5. These observations are consistent with the theoretical finding that the local buildup of optical energy is strongly dependent on the termination orientation.

\subsection*{Broadband frequency dependence of enhancement}
The previous section studied the localized field enhancement for a particular laser frequency of 200.0 THz within the surface mode gap. We next investigate the enhancement in a broad frequency range. Fig.~\ref{fig4} shows a map of intensity as a function laser frequency and position along the waveguide. Each horizontal line is obtained in a similar way as the green integrated intensity line of Fig.~\ref{fig3}. Three different terminations with varying dislocations are compared: two with a zigzag symmetry and one with an armchair symmetry. This comparison confirms that enhancement of field intensity only occurs at a zigzag termination and not at an armchair termination. We observe that the enhancement is present over $\sim$6~THz, and is the largest for frequencies within the surface mode gap. Its magnitude, which is the ratio between the maximum intensity of light at the termination and the average intensity of forward propagating light in the waveguide, varies with frequency, and reaches a maximum just below the upper limit of the gap (white dashed line in Fig.~\ref{fig4}A). Supplementary Materials Fig. S4 depicts the spectral shape of this broadband peak. 

A possible interpretation of both the frequency dependence and the sharply defined spatial extent of the localized enhancement could be sought in the involvement of evanescent states of the VPC-PhC interface at the termination. Within the surface mode gap, the incident edge state cannot excite propagating states with a real wavevector along the VPC-PhC interfaces. However, a continuum of evanescent states with an imaginary wavevector along the interfaces can be expected to exist in the  gap~\cite{Joannopoulos2008}, which could be excited given appropriate wavefunction overlap with the incident waveguide field. These states cannot propagate, but could radiate into either free space or the backward propagating mode. In this interpretation, the suppressed scattering could be thought of as being assisted by compatible evanescent states at the interfaces, confined to the termination, that contribute to the buildup of local optical intensity there. As the overlap of incident and evanescent states would depend on the intricate details of their local fields, which are generally frequency dependent, their excitation amplitude can therefore vary slowly with frequency. In principle, evanescent modes would also exist at the armchair terminated waveguide with an overlap that may be significantly different from a zigzag terminated waveguide. The absence of any localization suggests that the intervalley scattering is much stronger and the edge states instead couple directly to the backward propagating mode. The significant difference in localization at this different symmetry hints that the coupling to evanescent states is negligible. This is indeed a remarkable distinction in scattering behaviour between the zigzag and armchair terminated waveguides. We note that this possible viewpoint of storage in local evanescent states does not provide an alternative explanation to the general effect of localization at a topological waveguide termination that we report, but it could provide a route to understanding the details of spatial and spectral extent of the intensity enhancement in future work.

We note that localization of optical energy is also observed experimentally at the zigzag termination for frequencies slightly above the upper gap edge. Within the same viewpoint, we note that for those frequencies upper surface modes could be excited as long as there is finite overlap with the wavefunction of the edge state. As such states will have low group velocity near the band edge, they are expected to contribute to local field enhancement. For the lower surface modes the overlap may be insufficient, as no enhancement is observed here. Only in the case of sufficient wave function overlap, and for the modes with a group velocity significantly lower than the group velocity of the edge state, modes can become confined at the termination and in the output ports. In some cases this results in an intensity in the output port that is higher than inside the topological waveguide, as seen in Fig.~\ref{fig2}A and Fig.~\ref{fig2}B. The magnitude of intensity in the output port however depends on how much light scatters out-of-plane or reflects back to the topological waveguide. Supplementary Materials Fig. S3 shows the optical energy at an armchair termination and shows no high levels of intensity in the output ports. Light is simply allowed to backscatter, resulting in a reduced amount of energy flowing to the output ports. The heterogeneous distribution of energy at distinct terminations further underscores the pivotal role of the conservation of the valley degree of freedom.

\section*{Discussion}
Using valley photonic crystals as a topological photonic platform, we demonstrate that light can be significantly localized at a termination of a reciprocal topological waveguide. Only for terminations that approximately conserve the valley DOF, reflection is suppressed strongly enough to result in localization. We observe a broadband peak of the optical intensity in the frequency range of the surface mode gap at a zigzag termination of the waveguide. For frequencies above the surface mode band gap, less prominent localization is attributed to the wavefunction overlap between the edge state and the upper surface modes with a low group velocity. Measurements of the local intensity at an armchair termination show no enhancement at all, confirming the role of the symmetry-protected topology in the enhancement mechanism. We provide a theoretical description explaining the origin of the suppressed backscattering and also confirm the experimentally observed light enhancement using simulations. 

This experimental observation of broadband light enhancement at a termination of a topological photonic waveguide is an example of a broader class of systems that could facilitate field enhancement due to suppressed backscattering~\cite{Mann2021}. Topologically non-trivial waveguides with a termination protected by lattice symmetry pave the way for the development of novel devices exploiting the potential of enhanced light matter interactions. Further research should explore the possible magnitude of broadband enhancement and the role of local design features, such as the dislocation and the choice of the terminating crystal. More broadly, the demonstrated paradigm of optical energy storage and localization, in combination with robust guiding and manipulation of light, present opportunities to on-chip photonic technology.

\section*{Materials and Methods}
\subsection*{Device fabrication}
We fabricate a PhC slab on a SOI platform with a 220 nm thick silicon layer on a 3 \textmu m  buried oxide layer. First, a positive electron-beam resist of thickness 240 nm  (AR-P 6200.09) is spin-coated. Then, the PhC design is patterned into the resist using electron-beam lithography on a Raith Voyager with 50 kV  beam exposure. The electron-beam resist is developed in pentyl acetate/O-xylene/MIBK:IPA(9:1)/isopropanol, and the SOI chip subsequently undergoes reactive-ion etching in HBr and O\textsubscript{2}. 

Next, a Suss MABA6 Mask Aligner patterns the photosensitive resist AZ1518 to define a selective wet-etching window on the PhC. After development with AZ400K:H\textsubscript{2}O, the buried-oxide layer is partially removed by wet etching in a 1:2 solution of 40\% hydrofluoric acid and deionized water for 20 min. The sample is then subjected to critical point drying to obtain free-standing PhC membranes \cite{Reardon2012}. Finally, to enable incoupling from the side, the SOI chip is cleaved on the end facets of the Si-ridge waveguides that extend to the PhCs. The Si-ridge waveguides are tapered down to support only the fundamental TE mode at the VPCs end facet. The ridge waveguides adiabatically transition to the VPC waveguide inside the lattice, enabling better index matching for efficient incoupling \cite{Shalaev2019}.

To account for fabrication imperfections that shift the operating frequencies of the devices, we fabricate triangular PhC lattices with three different lattice constants a$_0$ = [449 nm, 494 nm, 539 nm]. In the VPCs, a rhombic unit cell contains two triangular holes of side lengths $s_1$  = 0.65a$_0$ and  $s_2$  = 0.38a$_0$, respectively. A domain wall is created along VPC1 and VPC2 by applying a parity operation along the spatial y-coordinate. In the photonic bandgap material, a rhombic unit cell contains a single circular hole of radius r = 0.27a$_0$.

\subsection*{Near-field microscopy setup}
The near-field probe is fabricated by heated pulling of a single mode optical silica fiber using a P-2000 micropipette puller. Subsequently, the tips are etched in buffered oxide 7:1 for 20 minutes to ensure a wide opening angle of the aperture. The tips are coated homogeneously using a Temescal FC-20349 evaporator with first $\sim$1.5 nm chromium for improved adhesion and then $\sim$150 nm aluminium. The aluminium coating protects against radiative light loss at significant distances from the tip. The aluminium at the tip is then partly milled away by focused ion beam etching (Helios G4Cx) to create an aperture of 194 nm. 

Probes are attached to one prong of a Farnell AB38T quartz tuning fork with a resonance of 32.768 kHz. The tuning fork has a Q-factor of $\sim$700 and is soldered to a custom made chip that drives the tuning fork, and reads out its oscillating frequency. The chip is then connected to a piezo dither block which rasters scan the probe above the photonic crystal while being controlled by a shear force feedback loop via a digital lock-in amplifier (Zurich Instruments HF2LI), keeping the tip at an expected height of $\sim$20 nm above the PhC surface.

\subsection*{Laser system}
The light source deployed in the experiment is a continuous wave tunable laser (Santec TSL-710), which is operable in the infrared frequency range [1480 nm - 1640 nm]. In a heterodyne detection scheme, the beam is split into a signal and a reference path by a polarizing beam splitter and the frequency of the light in the reference path is shifted by 60 kHz using two acousto-optic modulators (AOMs). The signal path is then coupled into the photonic crystal ridge waveguide using a high numerical aperture (NA = 0.8) microscope objective. A near-field probe picks up a fraction of the evanescent fields above the sample. Subsequently, signal and reference path are recombined, which forms a beating signal at 60 kHz facilitating the extraction of the electric field components including phase resolution with a high signal-to-noise ratio via a pair of lock-in amplifiers (Stanford Research Systems SR830).

\subsection*{Finite Element Frequency-Domain Simulation}
All numerical simulations of the photonic structure are performed using the finite element frequency-domain software COMSOL Multiphysics. Due to the high cost of 3D simulations on the computation resource, we simplify the simulation by leveraging the symmetry property of the structure. The suspended Si slab is mirror symmetric about the mid-height plane ($z=t/2$, where t is the slab thickness). Therefore, all electromagnetic modes can be classified as TE-polarized [$\mathbf{E}(z=t/2)\perp\hat{z}$] and TM-polarized [$\mathbf{E}(z=t/2)\parallel\hat{z}$] according to their field distributions at that plane \cite{Joannopoulos2008}. We focus on TE-polarized modes. By setting the mid-height plane of the slab as a perfect magnetic conductor, we simulate only half of the entire geometry, and only TE-polarized solutions are allowed. Above the slab ($\varepsilon_{slab}=12.11$), an air domain ($\varepsilon_{air}=1$) with a thickness of 3a$_0$ is included in the simulation domain. Except for the mid-height plane of the slab, all other boundaries are set using the scattering boundary condition.



\clearpage

\section*{Acknowledgments}

\subsection*{Acknowledgment}
The authors thank Sander Mann and Andrea Al\`{u} for inspirational discussions.

\subsection*{Funding:}
This work is part of the research programme of the Netherlands Organisation for Scientific Research (NWO). We acknowledge support from the European Research Council (ERC) Starting Grant no. 759644-TOPP.

\subsection*{Author contributions:} G.S. and E.V. conceived the experiment, Y.L. and R.B. defined and planned the experiments with input from all authors, D.M. conducted and analyzed the optical experiments, Y.L. performed the simulations, R.B. fabricated the samples, S.A. provided critical feedback and experimental support and L.K., G.S., and E.V. supervised the project. D.M. and Y.L. wrote the manuscript with contributions from all authors.
\subsection*{Competing interests:} The authors declare that they have no competing interests.
\subsection*{Data and materials availability: } All data needed to evaluate the conclusions in the paper are present in the paper and/or the Supplementary Materials. Additional data related to this paper may be requested from the authors.

\clearpage
\section{Figures}

\begin{figure}[h!]
    \centering
    \includegraphics[width=\textwidth]{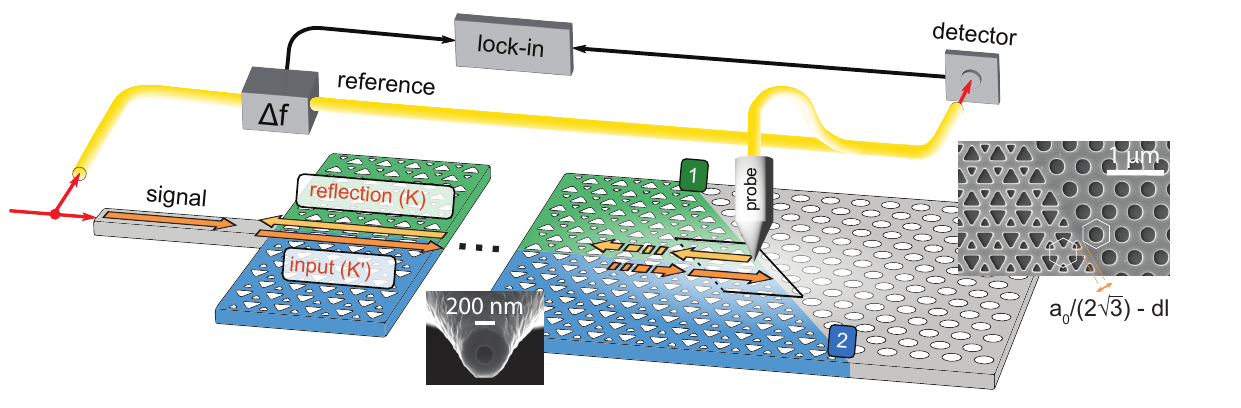}
    \captionsetup{labelformat=empty}
    \caption{}
    \captionsetup{labelformat=default}
    \label{fig1}
\end{figure}

\textbf{Fig. 1. Schematic of the experiment.} High-level diagram of the near-field scanning optical microscope setup and the topological waveguide terminated by a trivial PhC. The colors (green and blue) depict two mirrored
non-trivial photonic crystals and the trivial photonic crystal (grey) at the termination. Orange arrows with different shades depict forward and backward propagating edge states in valleys K’ and K. Infrared light enters via a silicon ridge waveguide from the side of the sample and propagates robustly in the topological waveguide until it scatters at the termination. Light may couple into the output ports (1) and (2). A near-field probe raster scans above the surface and detects a fraction of the components of the in-plane electric field. The detected signal is recombined with the reference branch which is given a 60 kHz offset in frequency resulting in a beating signal that is read out by lock-in amplifiers. Insets show SEM images of a near-field probe and a symmetry protected zigzag termination. The distance between the VPCs and the trivial PhC depends on the dislocation parameter and affects the frequency range at which surface modes appear in the photonic dispersion diagram.

\clearpage

\begin{figure}[h!]
    \centering\includegraphics[width=\textwidth]{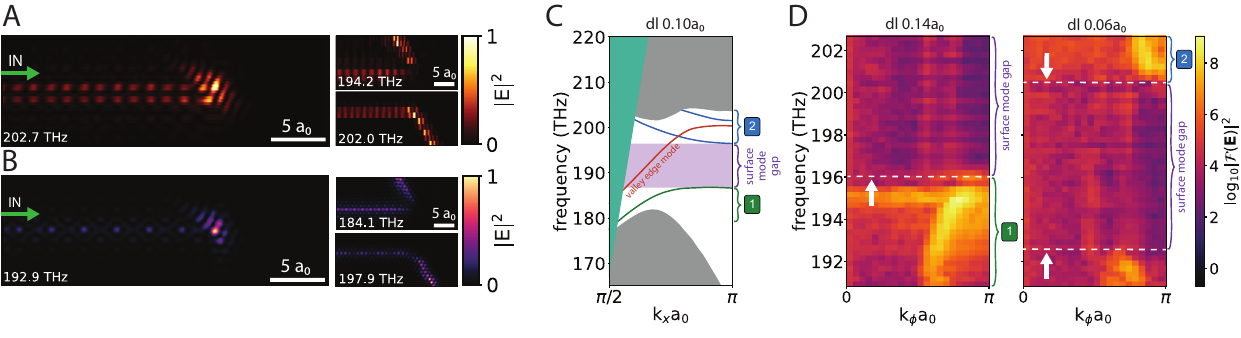}
    \captionsetup{labelformat=empty}
    \caption{}
    \captionsetup{labelformat=default}
    \label{fig2}
\end{figure}

\textbf{Fig. 2. Optical energy enhancement at the zigzag termination.} \textbf{(A)} Experimental measurement of the near-field in-plane electric field at a dislocation of 0.14a$_0$, normalized to the maximum of the scan at corresponding frequency (bottom left corner). Optical energy fed by the valley topological waveguide localizes at the termination for frequencies in the surface mode gap. Panels on the right show that for frequencies outside the surface mode gap the valley edge mode couples to the trivial surface modes resulting in propagation along the terminating interface. The upper (lower) panel depicts this propagation at a dislocation of 0.14a$_0$ (0.06a$_0$). \textbf{(B)} Near-field in-plane electric field at a dislocation of 0.10a$_0$ in simulation. \textbf{(C)} Simulation of the photonic band diagram for a dislocation of 0.10a$_0$ showing the surface mode gap in between the surface modes (green and blue lines). The valley edge state (orange line) spans the entire surface mode gap, shown here in the vicinity of valley K' $\equiv 2\pi/3$. Bulk modes are located in the solid grey areas. Indicated by the solid green area is the region above the light line, where radiative losses occur. \textbf{(D)} Photonic band diagram in experiment, for a dislocation of 0.14a$_0$ and 0.06a$_0$ at the VPC-PhC interface. For a larger dislocation the surface mode bands shift up in frequency due to a stronger spatial confinement at the termination. Dashed lines indicate the limits of the surface mode gap. Bulk modes from the VPCs appear below 192.5 THz.

\clearpage

\begin{figure}[h!]
    \centering
    \includegraphics[width=\textwidth]{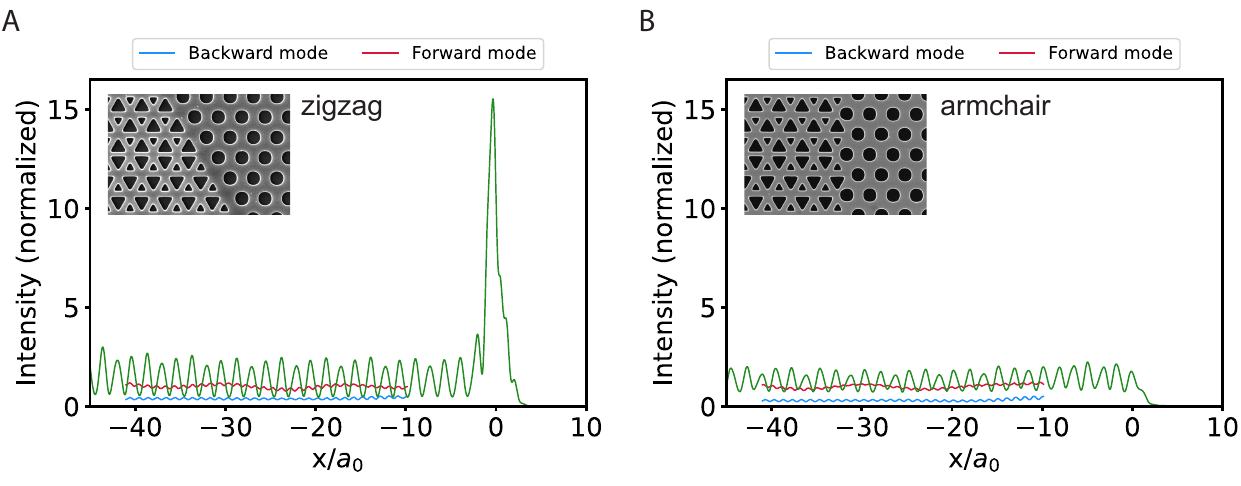}
    \captionsetup{labelformat=empty}
    \caption{}
    \captionsetup{labelformat=default}
    \label{fig3}
\end{figure}

\textbf{Fig. 3. Optical energy enhancement for the two terminations.} Experimental measurement of the intensity along the length of the topological waveguide, normalized to the intensity of the forward propagating mode. Light from the laser is coupled in at a frequency of 200.0 THz. The topological waveguide is terminated at position x/a$_0=0$. Energy localization only occurs at the zigzag termination, shown for dislocation 0.06a$_0$ \textbf{(A)} which nearly conserves the valley DOF and reduces backscattering. The spatial confinement of the enhancement peak is approximately 3a$_0$. The armchair termination, shown for dislocation 0.10a$_0$ \textbf{(B)} shows a uniform energy distribution in the waveguide and no enhancement. Segments of the forward- and backward-component of the electromagnetic wave retrieved using filters in k-space are shown in the middle of the topological waveguide. Insets: SEM images of zigzag and armchair VPC-PhC interfaces.

\clearpage

\begin{figure}[h!]
    \centering
    \includegraphics[width=\textwidth]{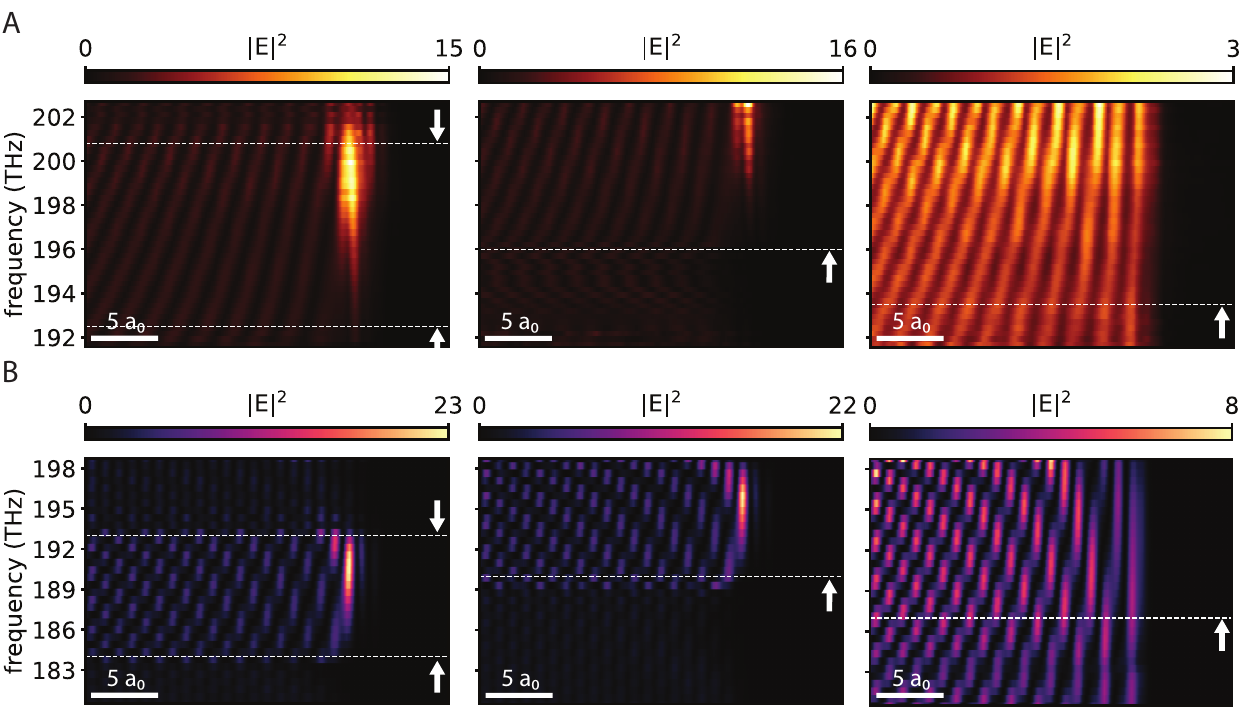}
    \captionsetup{labelformat=empty}
    \caption{}
    \captionsetup{labelformat=default}
    \label{fig4}
\end{figure}

\textbf{Fig. 4. Broadband field enhancement.} Experimental measurement \textbf{(A)} and simulation \textbf{(B)} of the field intensity inside the valley topological waveguide as functions of frequency. Plots are normalized by the average intensity of the forward propagating mode in the centre of the waveguide. From left to right: Zigzag termination with dislocation 0.06a$_0$, 0.14a$_0$ and armchair termination with dislocation 0.10a$_0$. At zigzag terminations, the optical energy is up to 15 times larger than in the middle of the waveguide; whereas at the armchair termination the optical energy is uniform along the entire waveguide. Horizontal dashed lines show the limits of the surface mode gap. The upper limit may appear out of the frequency range provided by the laser. 

\clearpage

\section*{Supplementary Materials}
Section S1. Design of the termination.\\
Fig. S1\\
Section S2. Extended near-field real space scans.\\
Fig. S2. \\
Section S3. Evaluation of the wave function overlap along zigzag and armchair terminations.\\
Table S1.\\
Section S4. Numerical recipe for separating the forward propagating and backward propagating components.\\
Fig. S3\\
Section S5. Frequency dependency of optical energy enhancement.\\
Fig. S4.\\
Fig. S5.\\
\clearpage

\subsection*{Section S1. Design of the termination}
This section describes how the design of the termination is established and explains the contribution of a dislocation value to the width between the VPCs and the trivial PhC. The termination of the topological waveguide described in this paper has two different symmetries, namely zigzag and armchair. The zigzag termination is oriented in the direction of the rhombic primitive cells of the VPC, at an angle of $\pi/3,2\pi/3$. Fig. S1A shows the primitive cells of the VPC lattice and the trivial lattice and illustrates the sizes of holes, which are inequivalent for the VPC lattice. The dimensions shown in Fig. S1A vary from the dimensions described in the main text in the Materials and Methods section because actual sizes after fabrication deviate from the initial design. Because the two VPCs are mirror inverted, two inequivalent VPC-PhC interfaces exist above and below the termination. At one interface the larger triangular hole of the VPC lies closest to a circular hole of the trivial PhC while for the other interface the smaller triangular hole lies closest. Fig. S1B schematically depicts this geometrical difference. Labels in the figure are similar to the labels of the output ports in the experimental setup shown in Fig.~\ref{fig1} in the main text. Light couples to an output port for a frequency range with equivalent label in Fig. 2D. The variation in geometry is the origin for the manifestation of two different surface modes at distinct frequencies. Both VPCs have the same hexagonal lattice structure with a triangular hole at each corner of the hexagon. The positions of these corners are equivalent for both VPCs and have equal distances to the hexagonal lattice of the trivial PhC. This distance amounts a$_0/\left(2\sqrt3\right)$ in the case of zero dislocation. We then subtract a dislocation term (dl) which slightly adjusts the distance between the lattices so that the VPC-PhC interface gap becomes narrower or broader. Adjusting the gap affects the spatial confinement of light and consequently the frequency at which a surface mode exists. We fabricate dislocations for the zigzag termination in steps of 0.02a$_0$, ranging from -0.06a$_0$ to 0.14a$_0$. For the armchair termination the geometry of the interface is equivalent below and above the termination. This equivalence results in the existence of only one surface mode, propagating in both ways along the terminating interface. This is shown in Fig. S3. At the armchair termination, corners of the hexagonal VPC-VPC lattice have equal distances to the corners of the trivial PhC hexagonal lattice everywhere along the terminating interface. Fig. S1B illustrates that this distance amounts 0.5a$_0$ and can be varied by subtracting a dislocation term (dl). For the armchair terminated waveguides we fabricate terminations with dislocations ranging from -0.20a$_0$ to 0.10a$_0$, in steps of 0.10a$_0$.

\clearpage

\begin{figure}[h!]
    \centering
    \includegraphics[width=\textwidth]{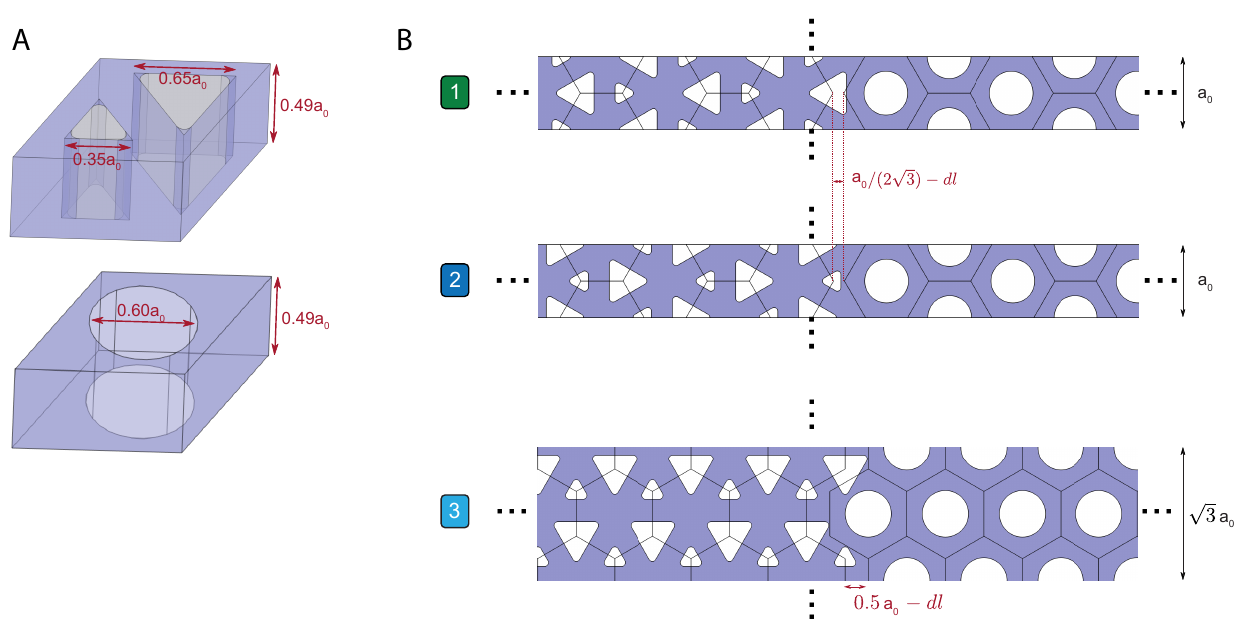}
\end{figure}

\textbf{Fig. S1. Lattice design.} Schematic representation of the lattice design. \textbf{(A)} The primitive cell of the VPC consists of two triangular holes of different size through a silicon slab. In the trivial PhC the primitive cell consists of only one circular hole. \textbf{(B)} Designs for the three different VPC-PhC interfaces for a single periodic length of the lattice. Labels 1 and 2 indicate designs with zigzag terminations whereas label 3 indicates an armchair termination. The VPCs and trivial PhC consist of a hexagonal lattice with equal lattice constant. The distance between the lattices depends on the symmetry of the termination and an assigned dislocation (dl) value. Larger dislocations correspond to less space between the lattices.

\clearpage

\subsection*{Section S2. Extended near-field real space scans}
The main text shows zoomed-in regions (27a$_0$ by 8a$_0$) of the near-field intensity close to the termination. Fig. S2 shows the original scans which are much larger (89a$_0$ by 67a$_0$) and almost capture the entire photonic crystal. The larger raster scans do not capture the position where light enters the photonic crystal because scattering is usually very high here. The field distribution of light in the topological waveguide and in the output ports is clearly visible. 
\clearpage
\begin{figure}[h!]
    \centering
    \includegraphics[width=\textwidth]{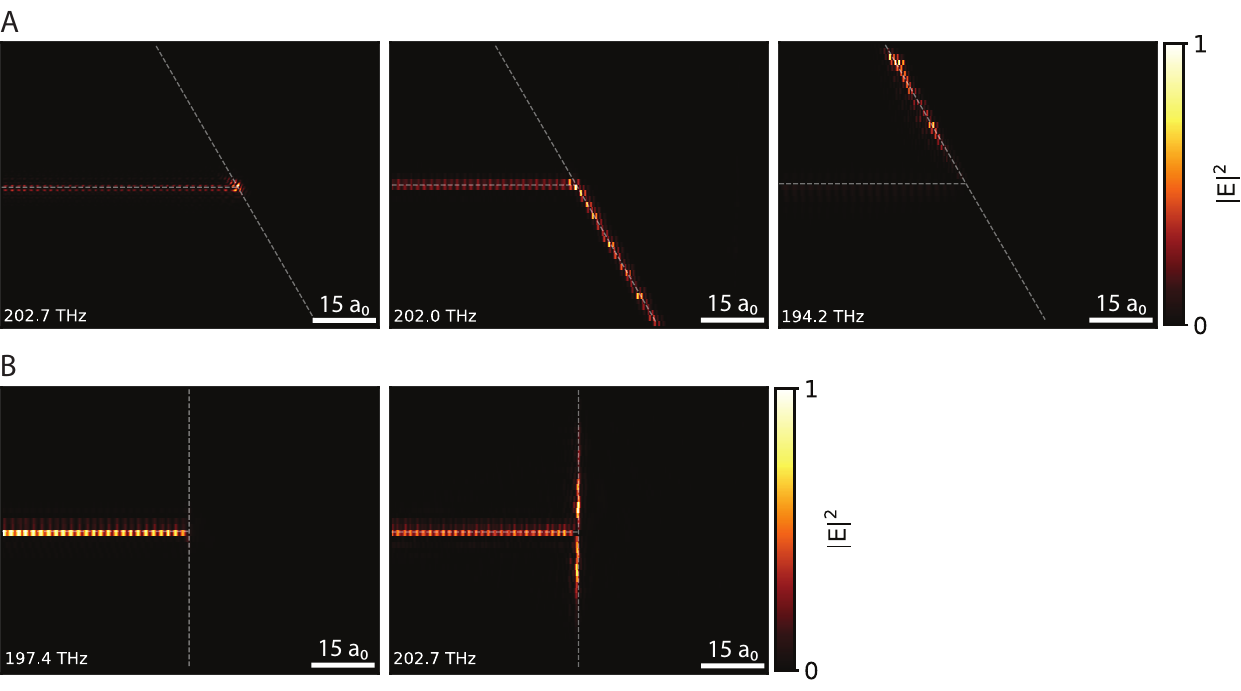}
\end{figure}

\textbf{Fig. S2. Extended real space scans.} Experimental measurements of the near-field in-plane electric field intensity, normalized to the maximum of the scan at corresponding frequency. \textbf{(A)} Optical energy fed by the valley topological waveguide localizes at the zigzag termination for a frequency in the surface mode gap. For a particular range of frequencies outside the surface mode gap the edge state couples to a surface mode and light propagates along the VPC-PhC interface to the edge of the PhC. \textbf{(B)} Optical energy fed by the valley topological waveguide is uniform and does not localize at the armchair termination for a frequency in the surface mode gap. For particular frequencies, surface modes exist to which the edge state can couple, resulting in light propagating in both ways along the VPC-PhC interface. Dashed grey lines illustrate interfaces of zigzag and armchair terminations.

\clearpage

\subsection*{Section S3. Evaluation of the wave function overlap along zigzag and armchair terminations}
(The following evaluation is rewritten based on the Supplementary Materials of Ref. \cite{Li2020}, which extends the derivation
in the Supplementary Materials of Ref. \cite{Ma2016})

In this section, we analyze how the quantum valley Hall (QVH) edge modes scatter when encountering zigzag and
armchair terminations. We demonstrate that the QVH edge mode weakly backscatters on a zigzag termination and
strongly backscatters on an armchair one. The backscattering coefficient, which is quantified as the overlap between
the two time-reversal conjugated modes (i.e., with opposite valley indices) $\Psi^{*}_{\mathbf{K'}} \Psi_{\mathbf{K}}$, shows a unique dependence on $\kappa$, the inverse of the edge mode’s transverse decay length.

The QVH edge mode can be written as $\Psi_\mathbf{K} = e^{i\mathbf{K} \cdot \mathbf{r}} u_\mathbf{K}(\mathbf{r})e^{-\kappa|y|}$. The overlap between the counterpropagating edge modes is 
$\Psi^*_{\mathbf{K'}}\Psi_\mathbf{K} = e^{i(\mathbf{K}-\mathbf{K'})\cdot \mathbf{r}}u^*_{\mathbf{K'}}(\mathbf{r})u_\mathbf{K}(\mathbf{r})e^{-2\kappa|y|}$,
where $\mathbf{K} = (4\pi/(3\text{a}_0), 0)$ and $\mathbf{K'} = (-4\pi/(3\text{a}_0), 0)$. $u_{K'(K)}$ are periodic functions with the same periodicity of the lattice. Therefore, $u_{\mathbf{K'}}u_\mathbf{K}$ also shares the same periodicity as the lattice, and the overlap can be written as,
\begin{equation}
\Psi^*_{\mathbf{K'}}\Psi_\mathbf{K} = e^{i\frac{8\pi}{3\text{a}_0}x}e^{-2\kappa|y|}\sum_{m,n}a_{m,n}e^{i(m\mathbf{b}_1 + n\mathbf{b}_2)\cdot \mathbf{r}}
\end{equation}
where $\mathbf{b}_{1,2}$ are the reciprocal lattice vectors, $\mathbf{b}_{1,2} = \frac{2\pi}{\text{a}_0}\left(1,\pm 1/\sqrt{3}\right)$. 

Then, we consider a termination with the same periodicity as the lattice. We can write the spatial distribution of the termination as a product of the components perpendicular and parallel to the termination direction, $A(\mathbf{r}_\perp)\left(\sum_l b_l e^{il \frac{2\pi}{\text{a}_0}\mathbf{r}_\parallel}\right)$.
The overlap of $\Psi^*_{\mathbf{K'}}\Psi_\mathbf{K}$ integrated along the terminating interface direction ($\mathbf{r}_\parallel$, also represented as the $\theta$-direction), is
\begin{equation}
\begin{gathered}
    A(\mathbf{r}_\perp)\int_{-\infty}^{\infty}dr_\parallel \sum_l b_l e^{il \frac{2\pi}{\text{a}_0}\mathbf{r}_\parallel}\psi^{*}_\mathbf{K'}\psi_{\mathbf{K}} \\
    = A(\mathbf{r}_\perp)\sum_{l,m,n}a_{m,n}b_l \int_{-\infty}^{\infty} dr_\parallel e^{il \frac{2\pi}{\text{a}_0}r_\parallel + i\frac{8\pi}{3\text{a}_0}x+i(m\mathbf{b}_1+n\mathbf{b}_2)\cdot \mathbf{r}_\parallel-2\kappa|y|}\\
    = A(\mathbf{r}_\perp)\sum_{l,m,n}a_{m,n}b_l \int_{-\infty}^{\infty} dr_\parallel e^{i \frac{2\pi}{\text{a}_0}[l+\left(\frac{4}{3}+m+n\right)\cos\theta+\sqrt{1/3}(m-n)\sin\theta]r_\parallel-2\kappa \sin\theta|r_\parallel|} \\
    = A(\mathbf{r}_\perp)\sum_{l,m,n}a_{m,n}b_l \int_{-\infty}^{0} dr_\parallel \left(e^{i \frac{2\pi}{\text{a}_0}[l+\left(\frac{4}{3}+m+n\right)\cos\theta+\sqrt{1/3}(m-n)\sin\theta]r_\parallel-2\kappa \sin\theta |r_\parallel|} \right. \\
    \quad \left. + e^{-i \frac{2\pi}{\text{a}_0}[l+\left(\frac{4}{3}+m+n\right)\cos\theta+\sqrt{1/3}(m-n)\sin\theta]r_\parallel-2\kappa \sin\theta |r_\parallel|)}\right) \\
    = A(\mathbf{r}_\perp)\sum_{l,m,n}a_{m,n}b_l \frac{4\kappa \sin\theta}{4\kappa^2(\sin\theta)^2 + \frac{4\pi^2}{\text{a}_0^2}\left[l+\left(\frac{4}{3}+m+n\right)\cos\theta+\sqrt{1/3}(m-n)\sin\theta\right]^2},
\end{gathered}
\end{equation}
where $\theta$ represents the direction of the termination: $\theta = 0, \pi/3, 2\pi/3$ correspond to the zigzag termination; $\theta = \pi/6, \pi/2, 5\pi/6$ correspond to the armchair termination. We shall see that it is the term $[l+\left(\frac{4}{3}+m+n\right)\cos\theta+\frac{1}{\sqrt{3}}(m-n)\sin\theta]$ in the denominator that determines the behavior of the overlap integral. For convenience, we represent this term with $c_{l,m,n}(\theta)$. Possible values of $c_{l,m,n}(\theta)$ for different terminations along the six different directions are listed in Table S1.

We notice that, with an armchair termination, $c_{l,m,n}(\theta)$ is allowed to attain a zero value while, with a zigzag termination, it is not. In the small $\kappa$ limit ($2\kappa \ll 2\pi/\text{a}_0$), the quantity \\ $4\kappa \sin\theta/\left[4\kappa^2(\sin\theta)^2 + \left(4\pi^2/\text{a}_0^2\right)(c_{l,m,n}(\theta))^2\right]$ is negligible as long as $c_{l,m,n}(\theta) \neq 0$.

According to Table S1, for zigzag terminations, the coefficient $c_{l,m,n}$ cannot be zero because a fractional number always remains. However, for armchair terminations, the coefficient $c_{l,m,n}$ can attain zero. With the specific combination of $\{l, m, n\}$ that makes $c_{l,m,n} = 0$, the quantity $4\kappa \sin\theta/\left[4\kappa^2(\sin\theta)^2 + \left(4\pi^2/\text{a}_0^2\right)(c_{l,m,n}(\theta))^2\right]$
diverges and makes the $\Psi^*_{\mathbf{K'}}\Psi_\mathbf{K}$ overlap diverge.

At last, we discuss the validity of the small $\kappa$ limit. $\kappa = \Delta/(2v)$, where $\Delta$ is the topological band gap width and $v$ is the slope of the Dirac cone before opening the band gap. For the microwave QVH photonic crystal in Refs. \cite{Li2020,Li2022}, $\Delta \approx 0.0398(2\pi c/\text{a}_0)$ and $v \approx 0.42c$, so $1/\kappa \approx 3.36\text{a}_0$, and the small $\kappa$ condition is satisfied.
\clearpage
\begin{table}[h!]
    \centering
    \begin{tabular}{|c|c|c|}
        \hline
        $\theta$ & termination shape & $c_{l,m,n}(\theta) \equiv l + \left(\frac{4}{3} + m + n\right) \cos \theta + \frac{1}{\sqrt{3}} (m - n) \sin \theta$ \\
        \hline
        $0$ & zigzag & $l + \frac{4}{3} + m + n$ \\
        \hline
        $\frac{\pi}{6}$ & armchair & $\frac{\sqrt{3}}{6} (4m + 2n + 4) + l$ \\
        \hline
        $\frac{\pi}{3}$ & zigzag & $m + l + \frac{2}{3}$ \\
        \hline
        $\frac{\pi}{2}$ & armchair & $l + \frac{1}{\sqrt{3}} (m - n)$ \\
        \hline
        $\frac{2\pi}{3}$ & zigzag & $-n + l - \frac{2}{3}$ \\
        \hline
        $\frac{5\pi}{6}$ & armchair & $\frac{\sqrt{3}}{6} (-2m - 4n - 4) + l$ \\
        \hline
    \end{tabular}
\end{table}
\textbf{Table S1. Values of $\mathbf{c_{l,m,n}(\theta)}$ for different terminations.}

\clearpage

\subsection*{Section S4. Numerical recipe for separating the forward propagating and backward propagating components}
This section describes how the forward- and backward propagating modes are separated in simulation and experiment. The intensity of the forward propagating mode is repeatedly used for normalization of the optical energy enhancement.
\subsubsection*{In simulation}
For convenience, we use $\mathbf{\psi}(\mathbf{r}, f)$ to represent the six-component electromagnetic field of the structure that is simulated using COMSOL at frequency $f$,
\begin{equation}
\psi(\mathbf{r}, f) \equiv \begin{bmatrix} E_x(\mathbf{r}, f), E_y(\mathbf{r}, f), E_z(\mathbf{r}, f), H_x(\mathbf{r}, f), H_y(\mathbf{r}, f), H_z(\mathbf{r}, f) \end{bmatrix}^T.
\end{equation}

In this section, we explain how to separate the forward ($+\hat{x}$)-propagating and the backward ($-\hat{x}$)-propagating components. First, we convert the field to the $\hat{x}$-directional momentum space (the $k_x$-space) using Fourier transform, 
\begin{equation}
\psi(k_x, y, z, f) \equiv \mathcal{F}\{\psi(x, y, z, f)\} = \int_{-\infty}^{\infty} \psi(x, y, z, f) e^{-i k_x x} \, dx.
\end{equation}
The integration is over the entire simulation domain.

$\psi(k_x, y, z, f)$ is symmetric about $k_x = 0$ (within the surface mode gap) because all the input energy is reflected, and the forward and backward components are identical in amplitude.

Due to the finiteness of the simulation domain, $|\psi(k_x, y, z, f)|$ approaches zero when $k_x \rightarrow \pm \infty$. We identify that the nonzero part of $\psi(k_x, y, z, f)$ dominates in the range $-2\pi/\text{a}_0 < k_x < 2\pi/\text{a}_0$. Therefore, the integration out of that range is discarded when calculating the inverse Fourier transform.

The $\mathbf{K'}$ valley ($k_x = (-2/3 + 2N)\pi/\text{a}_0$) corresponds to forward propagation; the $\mathbf{K}$ valley ($k_x = (2/3 + 2N)\pi/\text{a}_0$) corresponds to backward propagation, where $N \in \mathbb{Z}$. Therefore, we partition the range $-2\pi/\text{a}_0 < k_x < 2\pi/\text{a}_0$ into two parts,
\begin{align*}
    \text{toward right (TR)} & : -\pi/\text{a}_0 < k_x < 0, \quad \pi/\text{a}_0 < k_x < 2\pi/\text{a}_0, \\
    \text{toward left (TL)} & : -2\pi/\text{a}_0 < k_x < -\pi/\text{a}_0, \quad 0 < k_x < \pi/\text{a}_0.
\end{align*}

The forward propagating component, $\psi^{\text{TR}}(\mathbf{r}, f)$, and the backward propagating component, $\psi^{\text{TL}}(\mathbf{r}, f)$, are calculated by inverse-Fourier-transforming $\psi(k_x, y, z, f)$ in the two $k_x$ partitions, accordingly,
\begin{equation}
\begin{gathered}
    \psi^{\text{TR}}(x, y, z, f) = \frac{1}{2\pi} \int_{-\pi/\text{a}_0}^{0} \psi(k_x, y, z, f) e^{i k_x x} \, dk_x + \frac{1}{2\pi} \int_{\pi/\text{a}_0}^{2\pi/\text{a}_0} \psi(k_x, y, z, f) e^{i k_x x} \, dk_x, \\
    \psi^{\text{TL}}(x, y, z, f) = \frac{1}{2\pi} \int_{-2\pi/\text{a}_0}^{\pi/\text{a}_0} \psi(k_x, y, z, f) e^{i k_x x} \, dk_x + \frac{1}{2\pi} \int_{0}^{\pi/\text{a}_0} \psi(k_x, y, z, f) e^{i k_x x} \, dk_x.
\end{gathered}
\end{equation}
\\
\textbf{Normalization.} To compare the two structures, the waveguides with zigzag and armchair terminations, we normalize the electromagnetic field such that the input electromagnetic energy in the two setups is identical,
\begin{equation}
    avg(\text{S}_x^{TR, zig}) = avg(\text{S}_x^{TR, arm}),
\end{equation}
where $S_x = \text{Re}(E_y H_z^* - E_z H_y^*)/2$ is the time-averaged $\hat{x}$-directional Poynting vector. Here the average, $avg(\cdot)$, is taken over a waveguide segment away from the source and the termination.

\subsubsection*{In experiment}
A near-field probe picks up a fraction of the evanescent electric fields above the surface and retrieves all the spatial frequencies in the PhC at a sub-wavelength scale. Having access to these spatial frequencies, we can use a two-dimensional Fourier transform to obtain all wavevectors of light in the crystal for a single laser frequency. In two-dimensional reciprocal space we then observe all the (k$_x$, k$_y$) for which modes are found. An integration along k$_y$ then retrieves a distribution of intensity for wavevectors k$_x$, which is in the propagation direction of the topological waveguide. By executing the near-field raster scan for all frequencies that the laser provides and for each frequency retrieving the wavevectors k$_x$ we can reconstruct a dispersion diagram as shown in Fig. 2D. This diagram shows the dispersion of modes in valleys K and K' in the first Brillouin zone as well as the dispersion of all higher order Bloch harmonics. The slope of the dispersion indicates the sign of the group velocity $v_g=d\omega/dk$. A positive group velocity is indicative of a forward propagating mode while a negative group velocity refers to a backward propagating mode. We select an individual mode by selecting a specific interval of $k_x$ wavevectors. Subsequently we perform an inverse Fourier transform to real-space and obtain the real space electric field intensity in the topological waveguide.

\clearpage

\begin{figure}[h!]
    \centering
    \includegraphics[width=\textwidth]{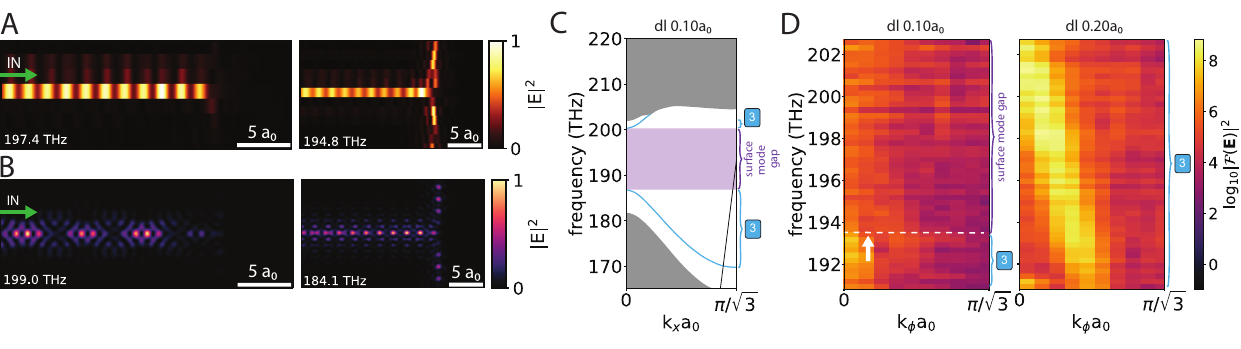}
\end{figure}

\textbf{Fig. S3. Optical energy at an armchair termination.} Experimental measurement \textbf{(A)} and simulation \textbf{(B)} of the near-field in-plane electric field at an armchair termination, normalized to the maximum of the scan at corresponding frequency  (bottom left corner). Optical energy fed by the valley topological waveguide does not localize at the termination. The right figure shows scattering of the valley edge mode to the trivial surface mode for frequencies outside the surface mode gap resulting in propagation along the terminating interface in both directions. \textbf{(C)} Simulation of the photonic band diagram for a dislocation of 0.10a$_0$ showing the surface mode gap and the dispersion of the surface mode (cyan line). Bulk modes are located in the solid grey area. Almost the entire diagram shows modes below the light line, meaning that radiative losses occur. \textbf{(D)} Experimental measurement of the photonic band diagram, for a dislocation of 0.10a$_0$ and 0.20a$_0$ at the VPC-PhC interface. For a larger dislocation the surface mode bands shift up in frequency due to a stronger spatial confinement at the termination. Dashed lines indicate the limits of the surface mode gap. Bulk modes from the VPCs appear below 192.5 THz.

\clearpage

\subsection*{Section S5. Frequency dependency of optical energy enhancement}

\begin{figure}[h!]
    \centering
    \includegraphics[width=\textwidth]{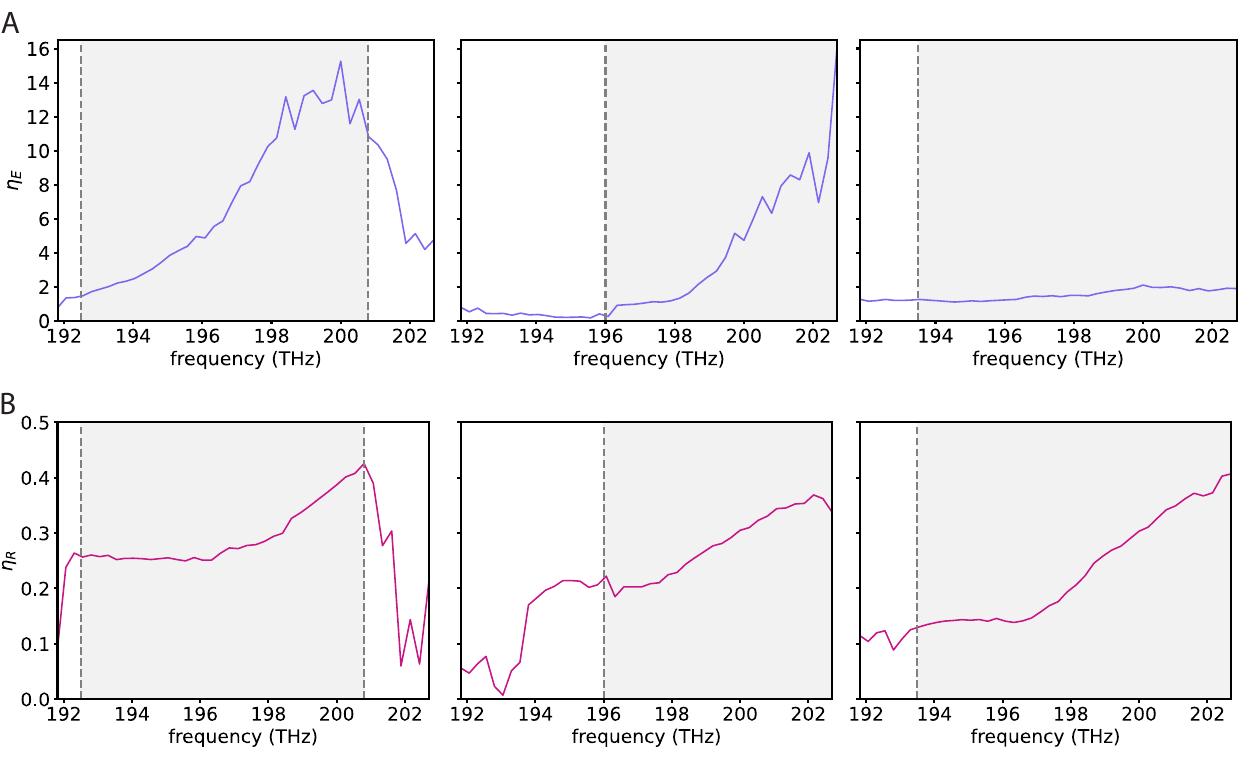}
\end{figure}

\textbf{Fig. S4. Enhancement and reflection coefficients.} Enhancement \textbf{(A)} and reflection \textbf{(B)} coefficients in experiment as a function of frequency for terminations with different geometries. From left to right: Zigzag termination with dislocation 0.06a$_0$, 0.14a$_0$ and armchair termination with dislocation 0.10a$_0$. The enhancement coefficient is given as the ratio between the maximum intensity at the termination and the intensity of the average forward propagating mode $\eta_E=I_{termination,\ max}/I_{s+}$. The broadband enhancement reaches it maximum just below the upper surface modes. The reflection coefficient is given as the ratio between intensity of the average backward and forward propagating mode $\eta_R=I_{s-}/I_{s+}$. Increasing reflection coefficients above $\sim$197~THz are accounted to the transition of the edge state past the light line. Below 197~THz, the edge state lies above the light line and can easily couple to free-space at the termination. As the group velocity of the edge state is comparable to that of the light line, the transition is broadband. The solid grey area depicts the broadband surface mode gap. 

\clearpage

\begin{figure}[h!]
    \centering
    \includegraphics[width=\textwidth]{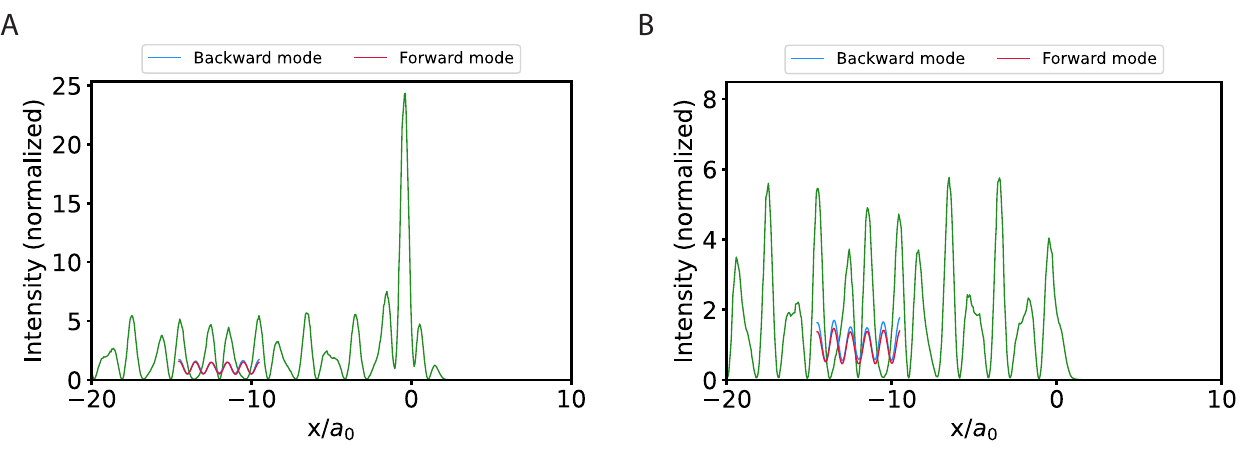}
\end{figure}

\textbf{Fig. S5. Simulation of the optical energy enhancement for the two terminations.} Optical energy along the length of the topological waveguide, normalized to the intensity of the forward propagating mode. Light from the source has a frequency of 191.0 THz. The
topological waveguide is terminated at position x/a$_0 = 0$. Energy localization only occurs
at the zigzag termination, shown for dislocation 0.06a$_0$ \textbf{(A)} which nearly conserves the valley DOF and reduces
backscattering. The armchair termination, shown for dislocation 0.10a$_0$ \textbf{(B)} shows a uniform energy distribution in the waveguide. The simulation domain is smaller than the experimental sample, due to the requirement for large computational resources for 3D simulations. Segments of the forward- and backward-component of the electromagnetic wave are shown in the
middle of the topological waveguide.

\end{document}